\documentclass[twocolumn,showpacs,preprintnumbers,amsmath,amssymb]{revtex4-1}

\usepackage{graphicx}
\usepackage{dcolumn}
\usepackage{bm}

\begin{document}

\title{Multi-Gaussian Modes of Diffusion in a Quenched Random Medium}
\author{Tapio Simula$^1$ and Mikko Stenlund$^{2,3}$}
\affiliation{$^1$School of Physics, Monash University, Victoria 3800, Australia \\ $^2$Courant Institute of Mathematical Sciences
New York, NY 10012, USA \\
$^3$Department of Mathematics and Statistics, P.O. Box 68, Fin-00014 University of Helsinki, Finland}
\pacs{02.50.-r, 05.40.-a, 66.30.-h}

\begin{abstract}
We have studied a model of a random walk in a quenched random environment. In addition to featuring anomalous diffusion and localization, for special regimes of disorder parameters the particle density decomposes into multi-Gaussian structure while its cumulative distribution is normal. We explain the observed fine structure of the density and point out its significance to experiments.

\end{abstract}

\maketitle

\section{Introduction}
Diffusion is ubiquitous in nature. The first theoretical description of diffusion processess such as Brownian motion of particles in colloidal suspensions was developed by Einstein and Smoluchovski \cite{Einstein1905a,Smoluchovski1906a}. Normal particle diffusion is characterized by the fact that the mean square displacement of the corpuscle grows linearly with time and the cumulative probability distribution is asymptotically that of a Gaussian density as dictated by the Central Limit Theorem (CLT). However, under special circumstances the situation may change dramatically, \textit{e.g.}, in Sinai's model the growth of the variance becomes logarithmically slow \cite{Sinai1982a}, and the particles may behave super- or subdiffusively or even become localized---a phenomenon predicted by Anderson for electron waves in disordered crystals \cite{Anderson1958a} and most recently observed with Bose--Einstein condensed matter waves in a controlled disorder \cite{Billy2008a,Roati2008a}.
 
Over the past few decades, the study of wave and particle propagation in random media has found a broad range of applications in diverse fields including medicine, optics, materials research, quantitative finance, and the biology of epidemics and genetic selection \cite{Gardiner2009a,Swishchuk2003a}. The approach to understanding random media has often been through random walks in random environments (RWRE) \cite{Kozlov1973a,Solomon1975a,Kesten1975a,Sinai1982a}. In these models the environment describes the local propagation laws, which model the local properties of the inhomogeneous medium in question \cite{Bolthausen2002a,Sznitman2004a,Zeitouni2006a}. In a \emph{quenched} (frozen) environment, the degree of stochastic freedom in the study of an RWRE is limited rendering the mathematical analysis of the problem challenging. In this setting, major advances have been obtained only very recently \cite{Goldsheid2007a,RassoulAgha2006a,Alili1999a}.

In this paper, we consider combined ballistic and diffusive particle dynamics \cite{Hellen2000a} in a spatially random but temporally fixed environment. Our model originates in the study of certain extended dynamical (\textit{i.e.}, deterministic) systems obtained by coupling together an infinite sequence of simple chaotic dynamical systems \cite{Koirat2009a}. With random initial conditions these can be reduced to RWREs in quenched environments by introducing a so-called Markov partition of the phase space. The model in question is also related to quenched compositions of chaotic maps \cite{AyyerLiveraniStenlund}.

\begin{figure}
\includegraphics[width=\columnwidth]{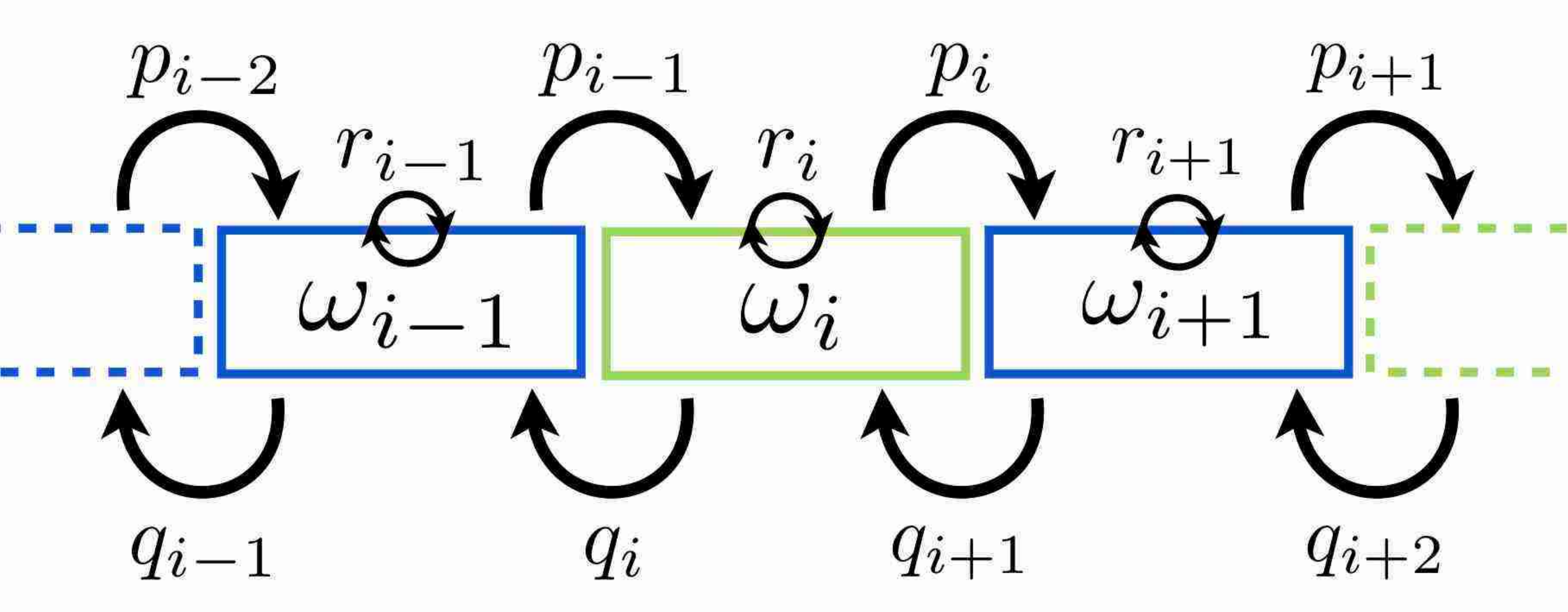}
\caption{(Color online) Schematic of our model of one-dimensional random walk in a quenched random environment. The labeled rectangles denote the sites of the discrete array and the arrows illustrate the site to site jump probabilities. Sites with the same blue or green color (dark or light gray) have the same fixed transition probabilities. The notation is explained in the text.}
\label{fig0}
\end{figure}

A natural question regarding such a stochastic process is whether the resulting particle distribution is asymptotically Gaussian. By virtue of Sinai's example \cite{Sinai1982a,Golosov1984a,Kesten1986a} one may not always expect an answer in the affirmative, even when the environment is not quenched. In our model the parameter space splits into subregions corresponding to different qualitative behavior; in some regions with ballistic behavior or parametric symmetries the cumulative distribution is normal, while in others it is not. Even in the former case the particle \emph{density} is not normal, and under certain circumstances features a multi-Gaussian structure---a striking phenomenon first observed in \cite{Koirat2009a}. Here we also explain the source of this observation.

\section{Model}
\subsection{Setup}
We consider one-dimensional RWRE as illustrated in Figure~\ref{fig0}. Within a single time step, from a given site there are prescribed probabilities of jumping into either of the two nearest neighbor sites. It is also permissible to remain in the same site with nonzero probability. Put more precisely, let $X_n\in \mathbf{Z}=\{\ldots,-1,0,1,\dots\}$ be the position of the particle at time $n\in\mathbf{N}=\{0,1,\dots\}$ and assume $X_0=0$. The particle jumps about the lattice $\mathbf{Z}$ according to Markov transition probabilities $p_{i,j}=\mathrm{P}(X_{n+1}=j\,|\,X_n=i)$. To describe these transition probabilities completely, we assign to each site $i$ a triplet $\omega_i=(q_i,r_i,p_i)$, where $q_i$, $r_i$ and $p_i$ are the probabilities of jumping to the left, staying put, and jumping to the right, respectively, if the current site is $i$. That is, $q_i=p_{i,i-1}$, $r_i=p_{i,i}$, $p_i=p_{i,i+1}$. In our model $q_i+r_i+p_i=1$, because the length of a jump cannot exceed one. We will also assume that $q_i>0$ and $p_i>0$. The collection  $\omega=(\ldots, \omega_{-1},\omega_0,\omega_1,\ldots)$ is called an environment.

Let $\{(q^\ell,r^\ell,p^\ell)\,:\,\ell\in\Lambda\}$ be the collection of all possible triplets that each $\omega_i$ may be equal to. Then, the triplets $\omega_i$ are drawn as i.i.d.\ trials from this collection with equal probability. Hence, for each site $i$, we have $\omega_i=(q^\ell,r^\ell,p^\ell)$ for some label $\ell\in\Lambda$, and this choice is independent of all the other sites. Once every $\omega_i$ has been randomly picked, it is frozen for good. An environment $\omega=(\ldots, \omega_{-1},\omega_0,\omega_1,\ldots)$ generated in this way is called a quenched random environment. 

For example, we can take $\Lambda=\{A,B\}$ if there are only two different kinds of triplets in the environment. Having chosen $\omega_i=(q^\ell,r^\ell,p^\ell)$ for a site $i$, where $\ell$ is either $A$ or $B$, we call interchangeably  $\omega_i$ and $\ell$ the label of the site. We also say that the site $i$ is of type $\ell$.

The sequence $(X_n)_{n\geq 0}$ of random variables is a Markov Chain and describes a random walk in a quenched random environment. We are interested in the probability distribution of $Z_n=(X_n-\mathrm{E}X_n)/\sqrt{\mathrm{Var}\,X_n}$ for typical environments $\omega$. In particular, CLT is said to hold, if $\lim_{n\to\infty} \mathrm{P}(Z_n\leq t) = \Phi(t)$ for all $t\in\mathbf{R}$. Here $\Phi(t)=\frac{1}{\sqrt{2\pi}}\int_{-\infty}^t e^{-s^2/2}\,ds$ is the cumulative distribution function of the standard normal distribution $\mathcal{N}(0,1)$. However, our focus is on the form of the density function $\mathrm{P}(X_n = k)$ itself, which admits surprising features.

\subsection{Implementation}
To compute the probability distribution of $X_n$ we have employed two complementary numerical methods. In the first method we calculate the exact probability distribution of $X_n$ by computing the $n$th power of the transition probability matrix of the Markov Chain. Since the lengths of the jumps are at most one and the walks start from $X_0=0$, we have $|X_n|\leq n$. Therefore, it suffices to compute powers of a finite matrix to find out the distribution of $X_n$ for any fixed $n$. Depending on the chosen parameters the distribution may in fact spread quite slowly with $n$ and a relatively small matrix can be used if $n$ is not too large. To study larger values of $n$, we use a Monte Carlo algorithm. Starting at $0$ and proceeding recursively, following a position $i=X_n$ the next position $X_{n+1}\in\{i-1,i,i+1\}$ is picked randomly according to the probabilities $(q_i,r_i,p_i)$. The procedure is repeated and averaged over many independent particle trajectories to obtain an approximation for the distribution of $X_n$. In the figures presented the former exact method has been applied.

\section{Preliminaries}
\subsection{Recurrence, transience, and ballisticity}
The walk is called recurrent if, with probability $1$ (w.p.\ $1$), it returns to $0$. In other words, at some later (random) time $n>0$, $X_n=0$. This clearly implies that, in fact, the walk returns to $0$ infinitely often. For our model it follows from \cite{Solomon1975a} that recurrence is equivalent to $\limsup_{n\to\infty}  X_n=\infty$ and $\liminf_{n\to\infty}  X_n=-\infty$ (w.p.\ $1$). The walk is called transient if it is not recurrent. In this case  one of only two things may happen \cite{Solomon1975a}. Either $\lim_{n\to\infty} X_n = -\infty$ (w.p.\ $1$) or  $\lim_{n\to\infty} X_n = \infty$ (w.p.\ $1$). In the former case the walk is transient to the left, in the latter it is transient to the right. While a transient walk does escape to infinity, it may do so extremely slowly, so that $\frac{X_n}{n}\to 0$. A transient walk is called ballistic if the limit $\lim_{n\to\infty} \frac{X_n}{n}$ exists and is nonzero. 

The quantity $\rho^\ell=\frac{q^\ell}{p^\ell}$ is called the bias of the label $\ell\in\Lambda$. 
The recurrence and transience properties of the walk can be conveniently described with the aid of the biases of the labels present in the environment; see below.

\subsection{Reversible state $\pi$}
There is a natural potential function $V$ associated with the environment $\omega$. Indeed, $\delta V_i=\log \frac{q_{i+1}}{p_{i}}$ can be thought of as a potential difference between the sites $i$ and $i+1$: if $\delta V_i>0$, the likelihood of jumping from $i$ to $i+1$ is less than vice versa, and the potential at $i+1$ should be higher than at $i$. To fix an arbitrary constant, we set $V_0=0$. We then define $V_{i+1}=V_i+\delta V_i$ for all $i$, such that $V_i = \sum_{0\leq j<i} \delta V_j$ for $i>0$ and $V_i = -\sum_{i\leq j<0} \delta V_j$ for $i<0$. A similar potential was considered in \cite{Sinai1982a}.

The Gibbs-like state $\pi$ defined by $\pi_i = e^{-V_i}$ satisfies the detailed balance condition $\pi_i p_{i,j} = \pi_j p_{j,i}$ ($|i-j|= 1$) and is therefore a reversible stationary state for the Markov Chain. This follows immediately from
\[
e^{-V_{i+1}} = e^{-(V_i + \delta V_i)} = e^{-V_i} \, \frac{p_{i,i+1}}{p_{i+1,i}}\, .
\]
Notice that $\pi$ cannot be normalized to a probability distribution, because in general $\sum_i \pi_i= \infty$. In fact, the potential $V_i$ may assume arbitrarily large negative values sufficiently far away from $i=0$, which makes $\pi_i$ unbounded. As we shall see, we can nevertheless learn about the distribution of the walk from the state $\pi$. It is particularly useful in situations in which the walk is recurrent.

\subsection{Stationary state $\mu$}
Now suppose the walk is transient to the right. Because of this property, it will visit every site $i$ finitely many times. Let $N_i$ be the total amount of time spent at site $i\geq 0$. Notice that, for every realization of the walk, there is precisely one more jump from site $i$ to $i+1$ than vice versa. This follows from transience to the right and the fact that the maximum jump length is one: every jump from $i+1$ to $i$ must sooner or later be followed by a jump from $i$ to $i+1$. On the other hand, the expected number of jumps from $i$ to $i+1$ is $\mathrm{E}(N_i)p_i$ and the expected number of jumps from $i+1$ to $i$ is $\mathrm{E}(N_{i+1})q_{i+1}$. Thus, we have the relation $\mathrm{E}(N_i)p_i = 1+\mathrm{E}(N_{i+1})q_{i+1}$.

The above reasoning motivates the following construction. Demand that $\mu_{i} = \frac{1}{p_i}(1+\mu_{i+1}q_{i+1})$ for every $i\in\mathbf{Z}$. Writing $\rho_i=\frac{q_i}{p_i}$ for brevity, the expression $\mu_i = \frac{1}{p_i}\bigl(1+\rho_{i+1}+\rho_{i+1}\rho_{i+2}+\rho_{i+1}\rho_{i+2}\rho_{i+3}+\dots\bigr)$ follows. Transience and ergodicity of the i.i.d.\ environment guarantee that the latter series converges. It is easily checked that the state $\mu$ so defined is a stationary state: the global balance condition $p_{i-1}\pi_{i-1}+r_i\pi_i+q_{i+1}\pi_{i+1} = \pi_i$ holds for every $i$.
\\

\section{Results}
\subsection{Anomalous probability density}

\begin{figure}
\includegraphics[width=\columnwidth]{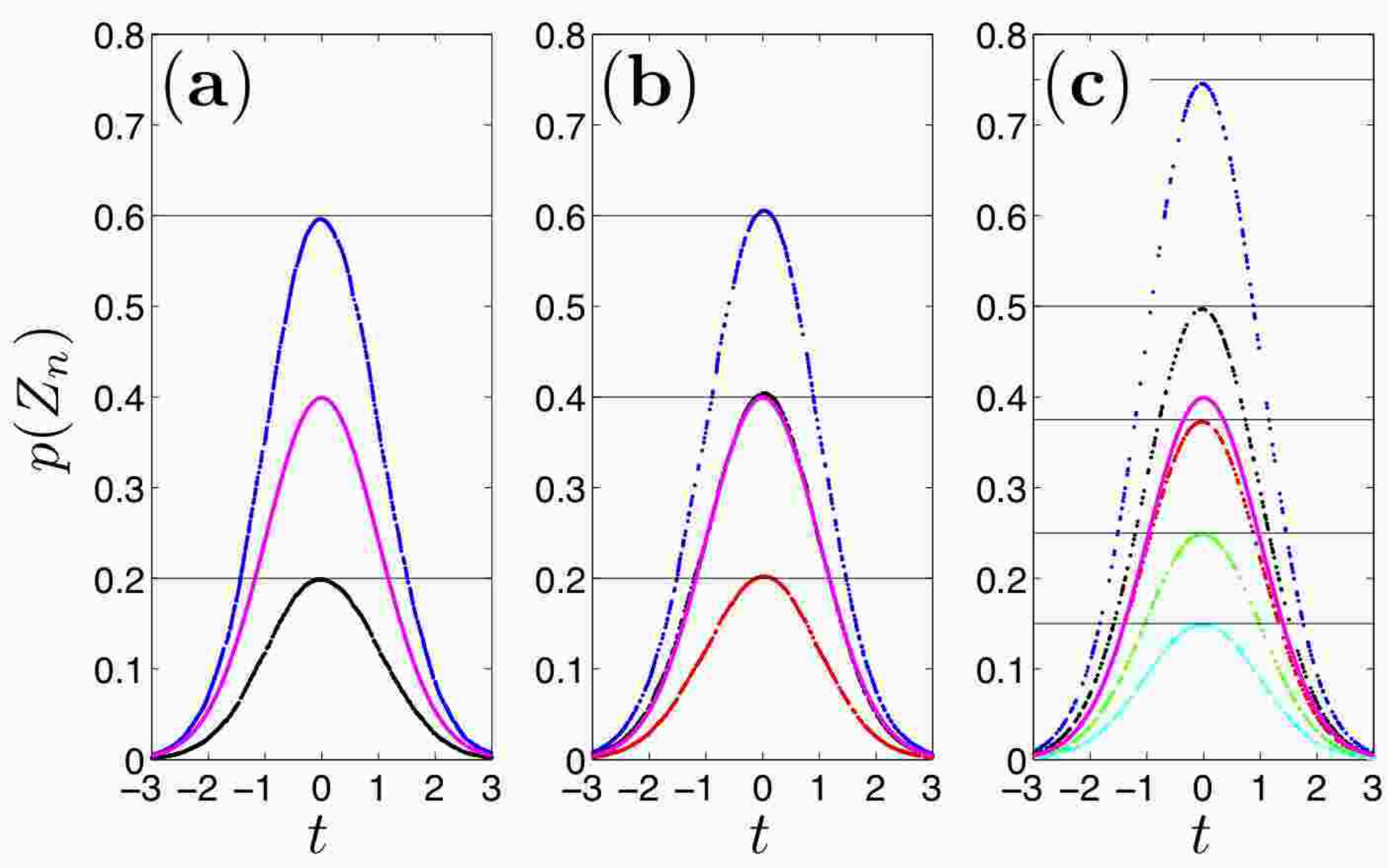}
\caption{(Color online) 
Probability densities of $Z_n$, where $n=10^5$, for a symmetric (a) two (blue and black dots) (b) three (blue, black and red dots) and (c) five (blue, black, red, green and cyan dots) label system, where the color (shade of gray) specifies the type of the endpoint label $\omega_{X_n}$. The transition probabilites are listed in the text. The ordinates of the horizontal lines, drawn to guide the eye, obey the same ratios as the right transition probabilities $p_\ell$ of the corresponding site types $\ell\in\Lambda$. The solid (magenta) curve in each frame is a Gaussian function with zero mean and a variance measured from the data. Each dotted curve coincides with the exact Gaussian when multiplied by a proper constant; see text.}
\label{fig1}
\end{figure}

In the  case where the bias $\rho^\ell$ has the same value for all labels $\ell\in\Lambda$, we find the probability distribution to decompose into $|\Lambda|$ distinct Gaussian functions, where $|\Lambda|$ is the number of elements in the set $\Lambda$.  The equal bias condition can be satisfied only if $r^\ell$ is different for every $\ell$, which means that staying put must be allowed if $|\Lambda|>1$. Figure~\ref{fig1} shows the scaled and centered probability densities of $X_n$ for the equal bias cases in which the sets of possible labels $\omega_i$ are
(a) {\tiny$\left\{(\frac{6}{10},\frac{1}{10},\frac{3}{10}),(\frac{2}{10},\frac{7}{10},\frac{1}{10})\right\}$}, 
(b) {\tiny$\left\{(\frac{2}{20}, \frac{14}{20},\frac{4}{20}),(\frac{3}{20},\frac{11}{20}, \frac{6}{20}),(\frac{6}{20},\frac{2}{20},\frac{12}{20})\right\}$} and 
(c) {\tiny$\left\{(\frac{2}{20},\frac{16}{20},\frac{2}{20}),(\frac{3}{20},\frac{14}{20},\frac{3}{20}),(\frac{4}{20},\frac{12}{20},\frac{4}{20}),(\frac{6}{20},\frac{8}{20},\frac{6}{20}),(\frac{10}{20},\frac{0}{20},\frac{10}{20})\right\}$}.
 In each frame, a different marker is assigned according to the environment label $\omega_{X_n}$ at the endpoint $X_n$. Joining the points of same marker type would result in perfect Gaussian curves which collapse on top of each other when multiplied by a constant. The ordinates of the horizontal lines, drawn to guide the eye, follow the same ratio as the right transition probabilities $p^\ell$ (equivalently the left transition probabilities $q^\ell$) of the labels on the corresponding curves. The latter curves themselves are similar; in each frame, multiplying any such curve with a certain constant results in perfect overlap with the solid Gaussian shown. Notice that in the equal bias case $\mu_i = \text{const}\cdot \frac{1}{p_i}$. Thus, at the local level, the particle distribution closely reflects the structure of the stationary state $\mu$. 
 
This picture is stable under variation of $n$: there are rapid oscillations from site to site, so the centered, scaled, and interpolated probability density does not converge to any function. Instead it has several smooth Gaussian ``components''. Altering the relative concentration of the site types only changes the concentration of points on each Gaussian. We emphasize that the multi-Gaussian structure only appears in the presence of the fixed environment. If one averages the distributions over many \emph{environments} (in addition to all possible trajectories which are already included),  the data collapses to a single Gaussian function. The multiple Gaussian curves persist as long as the biases $\rho^\ell$ are equal. From here on we focus on the two label case $\Lambda=\{A,B\}$. Increasing the number of site types does not affect our main conclusions.

\subsection{Analysis of observed densities}
The behavior of the probability density may be understood intuitively by considering the potential function $V_i$. Locally, particles tend to accumulate in the valleys of the potential function and correspondingly potential peaks repel particles. The bias $\rho_i$ yields the relative tendency of the site $i$ to transport particles to the left or to the right once they leave the site. Domain boundaries where neighboring sites tend to transport particles in the opposite directions correspond to the local extrema of the potential function. Knowledge of the structure of the potential, which is straightforward to obtain once the environment is specified, turns out to be enough to infer qualitative features of the density such as multi-Gaussianity without having to calculate or measure the density itself.

\begin{figure}
\includegraphics[width=\columnwidth]{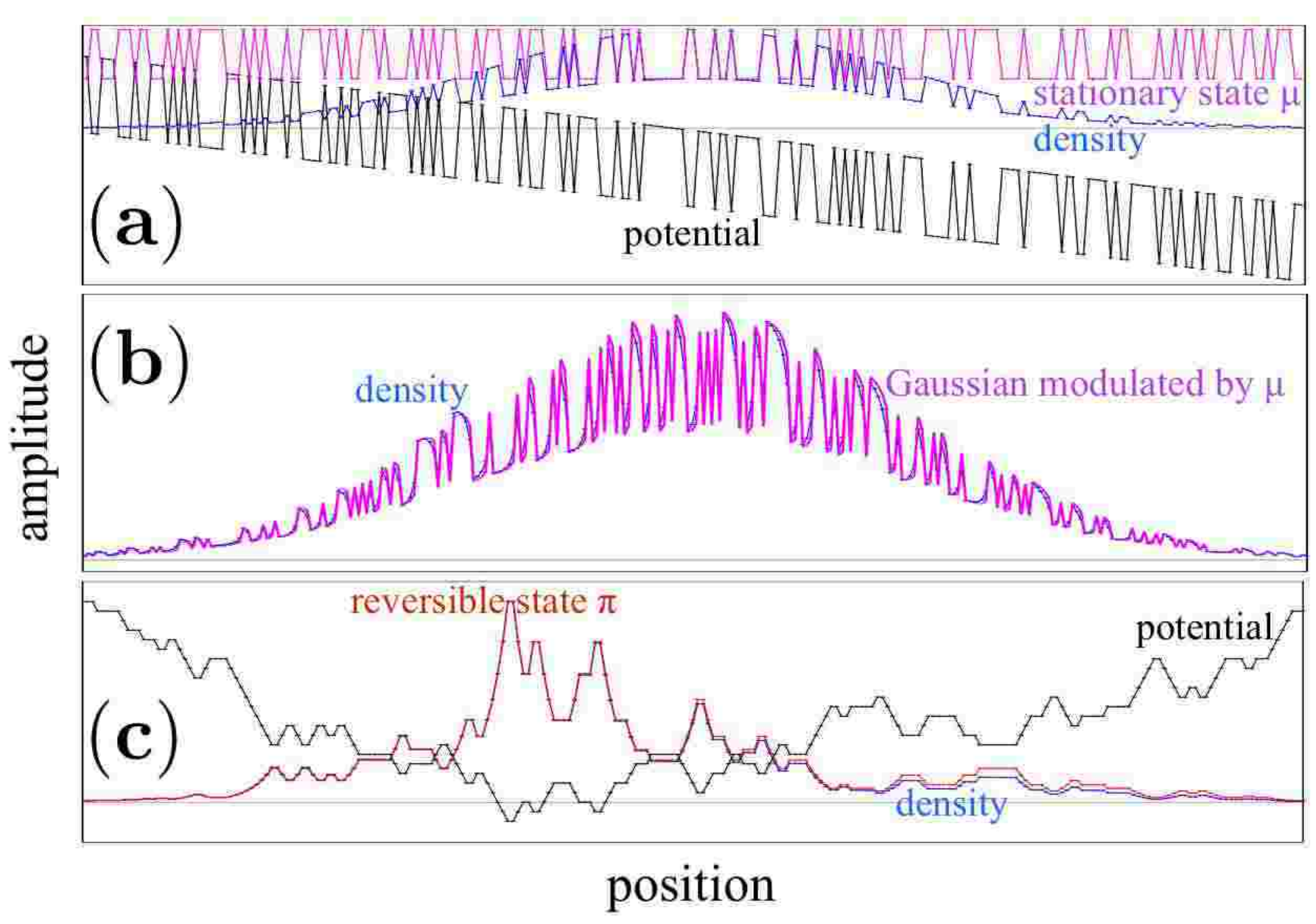}
\caption{(Color online) Probability densities for (a) $\rho^A=\rho^B\ne 1$, (b) $1>\rho^A>\rho^B$ and (c) $\rho^A\rho^B=1\ne\rho^A$. In (a) potential function, the stationary state $\mu$, and the density are plotted using different colors (shades of gray) and are tagged next to each of the curves.  In (b) the density and a Gaussian curve modulated by the stationary state $\mu$ closely follow each other (potential is not visible in this frame). Frame (c) shows the potential, density, and the reversible state $\pi$ as indicated next to each of the curves. Horizontal lines are drawn to guide the eye.}
\label{fig4}
\end{figure}

Figure~\ref{fig4}(a) shows a probability density together with the potential function for an equal bias case, $\rho^A=\rho^B\ne1$. In addition, the stationary state $\mu$ is plotted. The potential forms two linear potential levels resulting in the observed double Gaussian density. For general values of the biases, certain configurations of the environment labels yield clearly distinguishable linear potential levels. Correspondingly, the particle density has equally many Gaussian envelopes. For example, if the environment is made periodic, the potential (and hence also the density) always has multi-level structure also when $\rho^A\ne \rho^B$. This suggests to us that, whenever the CLT holds, the density in fact is a modulated Gaussian. The modulating factor at each site is a function of the environment but not of time.  Although the probability of sampling a perfectly periodic environment is 0, such environments may be physically important since many structures in nature exhibit approximate periodicity, including crystal structures in various solid state systems. Periodic environments can also be engineered on purpose. 

The evidence points to the stationary state $\mu$ being the modulator in the ballistic CLT regime. Indeed, the density in Figure~\ref{fig4}(a) is perfectly reproduced (less normalization) by multiplying $\mu$ with a Gaussian having the measured mean and variance. Further, Figure~\ref{fig4}(b) shows such modulated Gaussian which closely follows the actual density for a more generic transient case. The potential function is not visible in Fig.~\ref{fig4}(b) since it falls outside the scale shown in the figure. 

In Figure~\ref{fig4}(c), $\rho^A\rho^B=1\ne\rho^A$ and the particles become localized in the potential wells and make occasional quick journeys from one well to the next, which is analogous to tunneling through potential barriers. In this recurrent case, an excellent prediction for the density is given by the reversible state $\pi$ as illustrated in  Fig~\ref{fig4}(c). Notice, in particular, how the valleys and peaks in the potential are correlated with the maxima and minima, respectively, of the probability density.

\subsection{Phase diagram}
Let us collate a few known results relevant to our two label model, $\Lambda=\{A,B\}$. With the exception of recurrence, transience, and ballisticity, these results have been proved for the case in which staying still is forbidden ($r^\ell=0$). Recall  that $\rho^\ell=\frac{q^\ell}{p^\ell}$ is the bias of the site type $\ell = A,B$.

According to \cite{Solomon1975a,Zeitouni2006a}, if $\rho^A\rho^B=1$, the walk is recurrent. If $\rho^A\rho^B<1$, the walk is transient to the right. If in addition $\rho^A+\rho^B<2$, the walk is ballistic. By symmetry, left transience corresponds to $\rho^A\rho^B>1$. Below we will discuss recurrent and right transient cases.

\begin{figure}
\includegraphics[width=0.8\columnwidth]{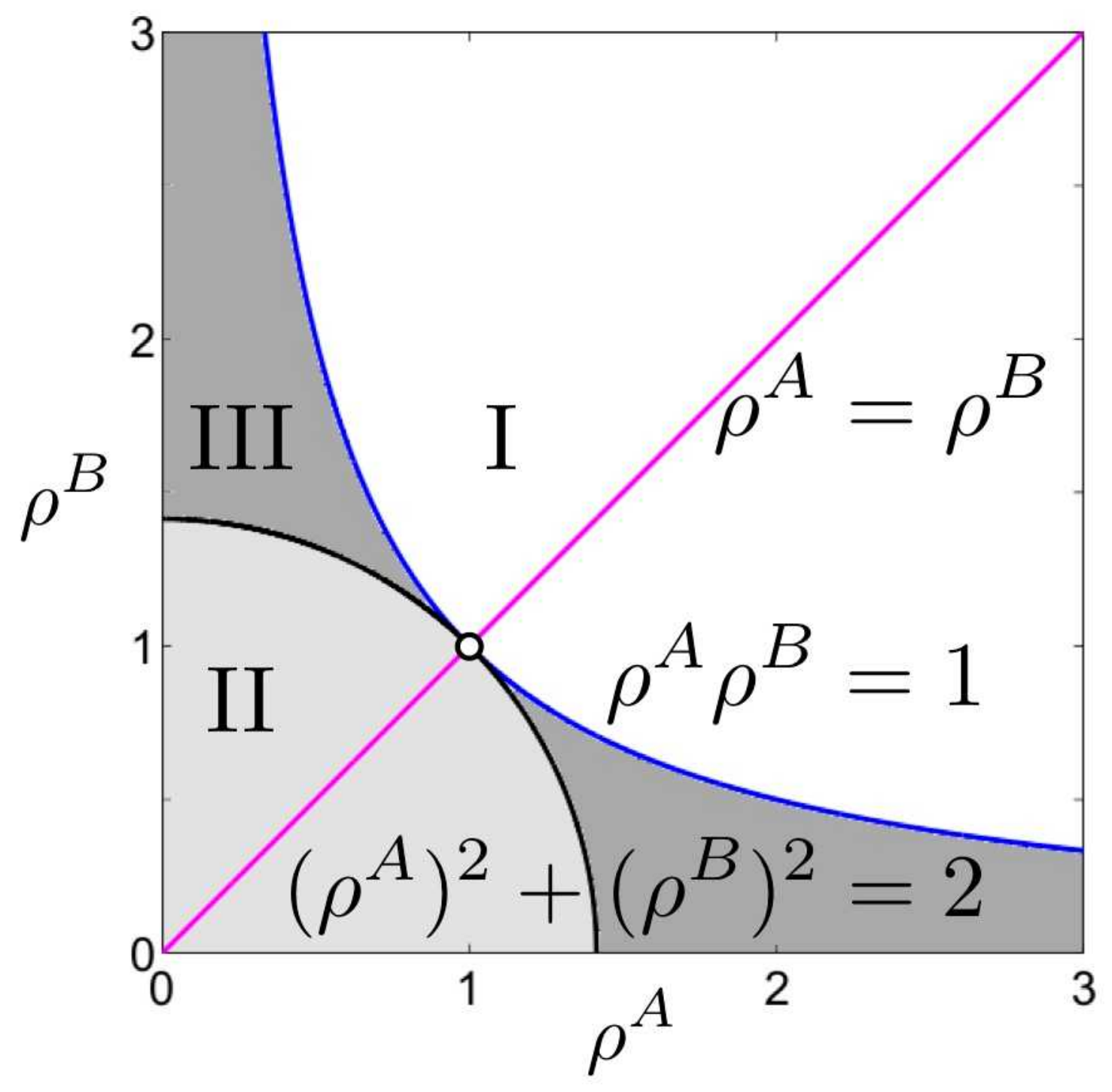}
\caption{(Color online) Schematic diagram displaying parameter regions exhibiting qualitatively different system behavior (assuming $r^\ell=0$).  On the blue (light gray) hyperbola $\rho^A\rho^B=1$ the walk is recurrent and shows Sinai diffusion, except precisely at $\rho^A = \rho^B = 1$ where CLT holds. In region (I) the walk is transient to the left. In regions (II) and (III) the walk is transient to the right. Inside the black (dark gray) circle $(\rho_A)^2+(\rho_B)^2 = 2$ the CLT holds. In the region (III) between the hyperbola and the circle, the CLT does not hold. When $r^\ell\neq 0$, double-Gaussian densities emerge on the magenta line $\rho^A=\rho^B$.}
\label{fig2}
\end{figure}

In the recurrent case, it has been shown \cite{Bolthausen2008a} that the CLT holds under the symmetry $\rho^\ell=1$ ($\ell=A,B$). The reason is that the stochastic process $(X_n)$ is a martingale due to the symmetry. In the absence of this symmetry the variance grows according to the Sinai law $(\log n)^4$. Sinai \cite{Sinai1982a} studied walks for which the probabilities $p_{i}=1-q_{i}$ ($r_i=0$) are i.i.d.\ random variables whose values are not quenched, meaning that the walks are averaged over all environments. His result is that,  in the recurrent case the variance of $X_n$ grows as $(\log n)^4$, which is astoundingly slow compared with the usual $n$. The limit distribution is not Gaussian: it has been computed explicitly by Kesten and Golosov \cite{Golosov1984a,Kesten1986a}. However, if one allows $r_i>0$ and imposes the additional inter-site symmetry $p_i = q_{i+1}$ for all $i$, which yields a flat potential ($\delta V_i = 0$), the usual CLT does hold \cite{Alexander1981a}.

Under the assumption $r^\ell=0$ and $\rho^A\rho^B<1$ (transience to the right), Goldsheid \cite{Goldsheid2007a} has shown that $(\rho^A)^{2}+(\rho^B)^{2}<2$ suffices for the CLT. Now, if $\rho^A> 1$, then $\rho^B\leq 1$ by transience, such that the equation $(\rho^A)^{s}+(\rho^B)^{s}=2$ has precisely one positive solution $s>0$. If $0<s<2$, the statistical behavior of $X_n$ is complicated and there is no limit distribution \cite{Peterson}. Independently of Goldsheid, Peterson \cite{Peterson} has also obtained the CLT, under stronger assumptions which include $s>2$. 

Figure~\ref{fig2} summarizes the above knowledge of the case $r^\ell=0$ ($\ell=A,B$) in the form of a schematic phase diagram of the system in terms of the biases $\rho^A$ and $\rho^B$. Superimposed is the double-Gaussian region we have observed when $r^\ell>0$.

\begin{figure}
\includegraphics[width=\columnwidth]{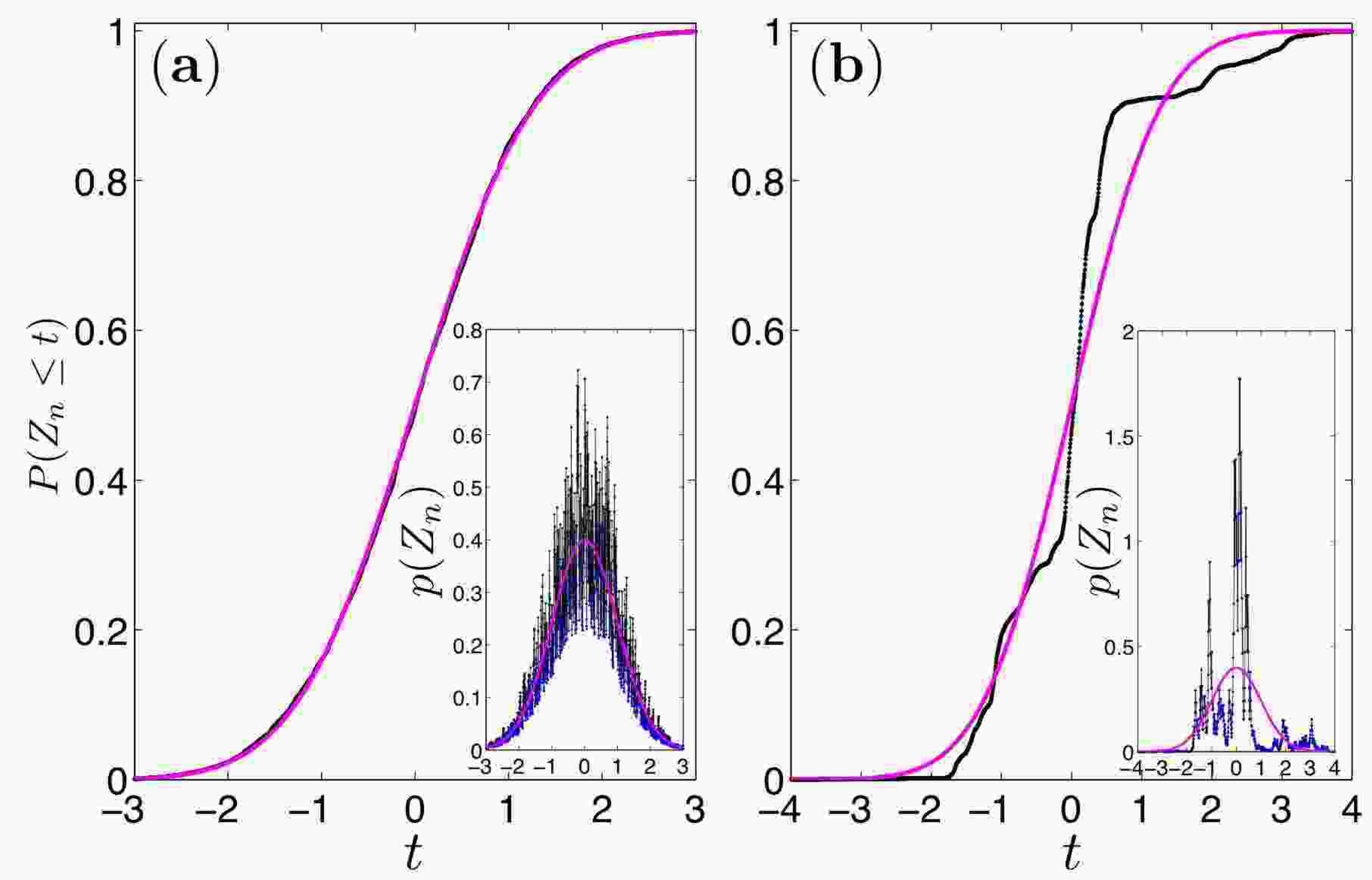}
\caption{(Color online) Cumulative distribution function for (a) $1>\rho^A>\rho^B$ and (b) $\rho^A\rho^B=1\ne\rho^A$. Insets show the corresponding probability densities where the solid curves are Gaussian functions with zero mean and whose variances correspond to the data. For both frames $n=10^5$.}
\label{fig3}
\end{figure}

For $\rho^A=\rho^B$, double-Gaussian density functions emerge. If $\rho^A \approx \rho^B$, both the probability density and the underlying potential exhibit a double structure. However, only for $\rho^A=\rho^B$ are the two densities strictly Gaussian. The larger the difference in the biases the larger the local fluctuations in the components of the probability density such that eventually the fluctuations grow to become of the order of the amplitude of the curves themselves and consequently the double Gaussian structure is washed away as illustrated in Figure~\ref{fig3}(a), where $\omega_i\in\left\{(\frac{4}{10},\frac{1}{10},\frac{5}{10}),(\frac{3}{10},\frac{0}{10},\frac{7}{10})\right\}$. When $\rho^A<1$ and $\rho^B\geq 1$, or vice versa, irregularity of the distribution even at the cumulative level is observed in simulations, as shown in Figure~\ref{fig3}(b), where $\omega_i\in\left\{(\frac{4}{10},\frac{1}{10},\frac{5}{10}),(\frac{5}{10},\frac{1}{10},\frac{4}{10})\right\}$. Whether this irregularity eventually disappears or persists forever depends on whether the parameters are in the CLT regime --- which  at least for $r^\ell=0$ is $(\rho^A)^{2}+(\rho^B)^{2}<2$ --- or not.

\subsection{Scaling of variance}
\begin{figure}
\includegraphics[width=0.8\columnwidth]{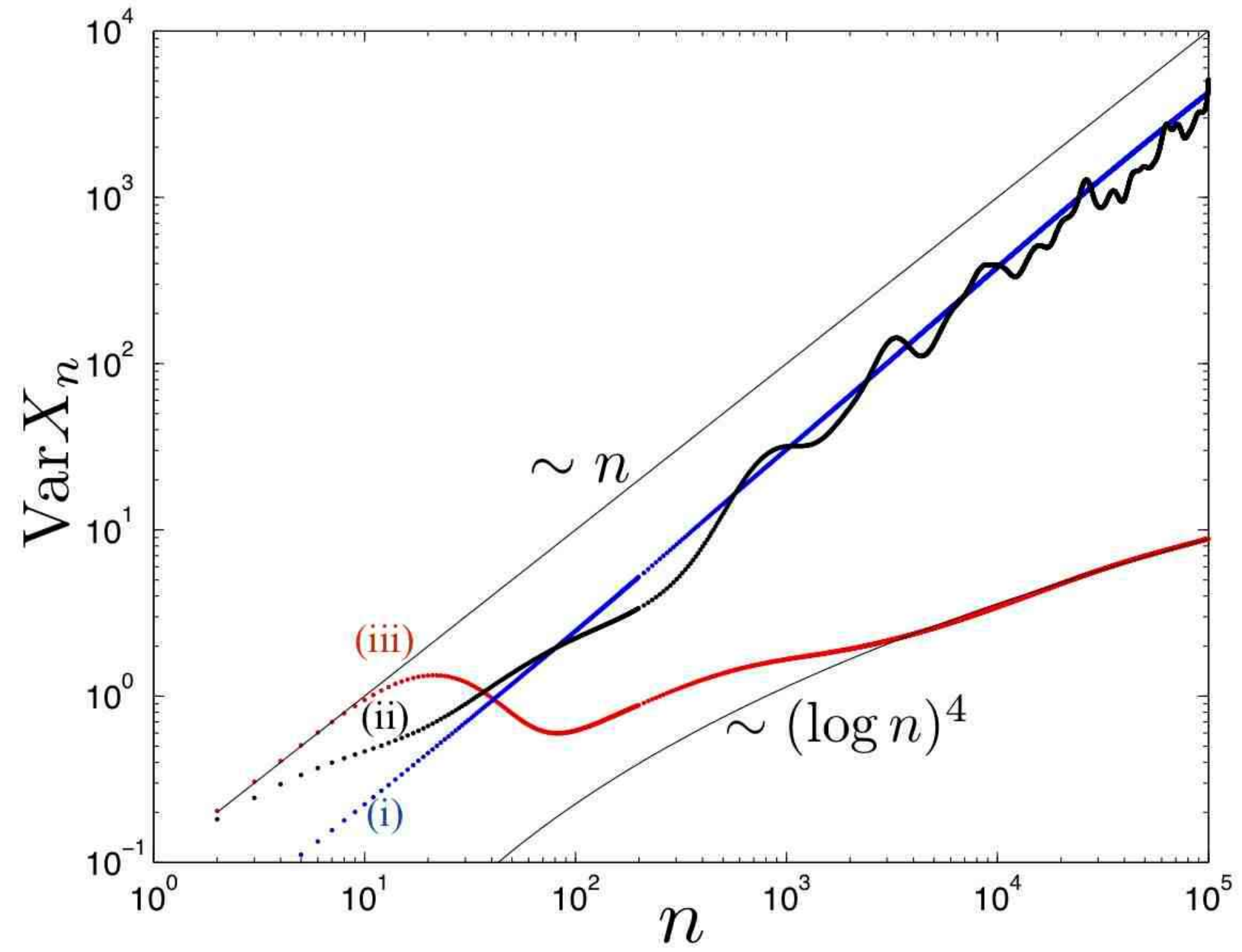}
\caption{(Color online) Variances as functions of time for 
(i) $\rho^A=\rho^B=1$,
(ii) $\rho^A=\rho^B\neq 1$, and
(iii) $\rho^A\rho^B=1\neq\rho^A $. The functional forms of the solid curves are marked in the figure. With nearly equal bias the variance grows linearly with time exhibiting normal diffusion. When at least one of the biases is different from unity the variance begins to fluctuate. In the strongly trapping case (opposite bias) particles become strongly subdiffusive and localize.}
\label{fig5}
\end{figure}

Figure~\ref{fig5} shows the variance of $X_n$ as a function of $n$ computed for three different parameter sets: (i) $\omega_i\in\left\{(\frac{10}{100},\frac{80}{100},\frac{10}{100}),(\frac{10}{900},\frac{880}{900},\frac{10}{900})\right\}$, (ii)  $\omega_i\in\left\{(\frac{1}{100},\frac{89}{100},\frac{10}{100}),(\frac{1}{900},\frac{889}{900},\frac{10}{900})\right\}$, and (iii) $\omega_i\in\left\{(\frac{10}{100},\frac{89}{100},\frac{1}{100}),(\frac{1}{900},\frac{889}{900},\frac{10}{900})\right\}$. For (i) the potential function has horizontal two-level structure and the variance grows linearly with $n$ exhibiting normal diffusion. 

In case (ii) the two-level potential is tilted inducing global drift to the system and the resulting variance shows prominent persistent oscillations. The system mimics normal diffusion on average but alternates between sub and superdiffusive behavior. This potentially transient behavior is understood to be caused by the probability mass propagating through focusing and defocusing regions of the fixed potential landscape. For practical purposes the fluctuations are quite prominent and should be significant for experiments. 

The third case (iii) demonstrates the effect of strongly trapping quenched environment whence the particles need to tunnel through arbitrarily high potential barriers and the system exhibits localization, as expected on the basis of the literature on the case $r^\ell=0$. By the same token, we typically observe localization of the density of the kind displayed in Figure~\ref{fig3} when $\rho^A\rho^B\approx 1$ (excluding the martingale case $\rho^A=\rho^B=1$), although by abstract results on the $r^\ell=0$ model this is likely to be a transient phenomenon if $(\rho^A)^{2}+(\rho^B)^{2}<2$ (region II in the phase diagram).

\section{Discussion}
We have investigated a model of a random walk in a quenched random environment. In the existing literature, particle distributions have been studied at a coarse diffusive scale. Indeed, the CLT gives asymptotic approximations of the probabilities $\mathrm{P}(a\sqrt n< X_n-\mathrm{E}X_n\leq b\sqrt n)$ for fixed pairs $a<b$ in the large $n$ limit. In our work we have investigated substantially finer details of the distribution by studying $\mathrm{P}(X_n=k)$, for fixed values of $k$.

It turns out that the latter local distribution has an unexpectedly rich, highly oscillatory, nature. This is true even in the CLT regime, in spite of the fact that the cumulative distribution obtained by integrating the particle density over large length scales reduces to a perfectly Gaussian cumulative function. Such prominent and persistent oscillations are significant to experiments in which information regarding the scaling limit of the system is not sufficient. In this paper we have obtained theoretical understanding of them.

For equal-bias situations, as well as for periodic environments with generic biases, the particle density is observed to break up into multiple Gaussian envelopes whose relative amplitudes are given by the ratio $p^\ell/p^{\ell'}$ for labels $\ell,\ell'\in\Lambda$. This ratio is found to be correctly predicted by the stationary state $\mu$. Generically, for transient cases the density can be predicted by multiplying $\mu$ with the Gaussian whose mean and variance match those of the observed distribution. In recurrent cases we find, as one would expect based on related literature, that the reversible state $\pi$ provides the most accurate prediction for the particle density.

For the sake of concreteness, we have used traditional terminology in that the walk represents a particle in a medium. This should not be construed as a restriction of applicability to other types of processes. Nor should it be mistaken to mean that the basic constituents, the sites (or `cells'), must be small let alone microscopic. We believe that the observations made here can be extended to a range of processes occurring in random or otherwise disordered environments. In this paper, we have studied an ensemble of noninteracting particles. In future, it will be interesting to investigate the universality, persistence and stability under weak particle interactions of the observed multi-Gaussian densities in both classical and quantum systems in the presence of quenched environments.

\begin{acknowledgments}
This work was supported by the Academy of Finland. We express our gratitude to Lasse Leskel\"a for enlightening discussions.
\end{acknowledgments}

\end{document}